\documentclass[aps,prb,twocolumn,floatfix,superscriptaddress]{revtex4}

 \usepackage{graphicx}
 \graphicspath{{figures/}}

 \usepackage{amsmath,amssymb}
 \usepackage{units,xspace}

 \newcommand{\micron}{\ensuremath{\unit{\mu m}}\xspace}
 \renewcommand{\vec}[1]{\ensuremath{{\boldsymbol #1}}\xspace}
 \newcommand{\uvec}[1]{\ensuremath{\hat{\boldsymbol #1}}\xspace}
 
 \newcommand{\avg}[1]{\left< #1 \right>}
 \newcommand{\abs}[1]{\left\vert #1 \right\vert}

\begin{document}

 \title{Characterizing and tracking
 single colloidal particles with video holographic microscopy}

 \author{Sang-Hyuk Lee}

 \author{Yohai Roichman}

 \affiliation{Department of Physics and Center for Soft Matter
   Research, New York University, New York, NY 10003}

 \author{Gi-Ra Yi}
 \affiliation{Korea Basic Research Institute, Seoul 136-713, Korea}

 \author{Shin-Hyun Kim}

 \author{Seung-Man Yang}

 \affiliation{National Creative Research Initiative Center for
 Integrated Optofluidic Systems and Department of Chemical and
 Biomolecular Engineering, Korea Advanced Institute of
 Science and Technology, Daejeon, 307-701 Korea}

 \author{Alfons van Blaaderen}

 \author{Peter van Oostrum}
 \affiliation{Soft Condensed Matter, Debye Institute, Utrecht University, 3508 TA Utrecht,
   The Netherlands}

 \author{David G. Grier}

 \affiliation{Department of Physics and Center for Soft Matter
   Research, New York University, New York, NY 10003}

 \date{\today}

 \begin{abstract}
 We use digital holographic microscopy 
 and Mie scattering theory to 
 simultaneously characterize and track individual colloidal particles.
 Each holographic snapshot provides enough information
 to measure a colloidal sphere's 
 radius and refractive index to within 1\%,
 and simultaneously to measure its three-dimensional
 position with nanometer in-plane precision and 10 nanometer
 axial resolution.
 \end{abstract}

 \maketitle

 In addition to their ubiquity in
 natural and industrial processes,
 colloidal particles
 have come to be prized as building blocks
 for photonic and optoelectronic devices,
 as probes for biological and macromolecular processes,
 and as model systems for fundamental studies of many-body
 physics.
 Many of these existing and emerging applications
 would benefit from more effective methods for tracking colloidal 
 particles' motions in three dimensions.
 Others require better ways to measure particles' sizes
 and to characterize their optical properties, particularly
 if these measurements
 can be performed on individual particles \emph{in situ}.

 This Article demonstrates that
 images obtained with
 in-line holographic microscopy \cite{sheng06,lee07}
 can be interpreted with Lorenz-Mie theory
 \cite{bohren83,barber90} to obtain
 exceptionally precise measurements of individual
 colloidal spheres' dimensions and optical properties
 \cite{ray91,denis06} while simultaneously 
 tracking their three dimensional motions with
 nanometer-scale spatial resolution at video rates \cite{moreno00}.
 This method works over the entire range of particle sizes and
 compositions for which Mie scattering theory applies, and requires
 only a single calibration of the optical train's magnification.
 Unlike other light scattering
 techniques for measuring particle size 
 \cite{xu02} or refractive index, 
 holographic particle analysis can be applied directly to
 individual particles in heterogeneous samples and
 also is compatible with
 scanned \cite{sasaki91}
 and holographic \cite{grier03} optical trapping.

 \begin{figure}[!t]
   \centering
   \includegraphics[width=0.65\columnwidth]{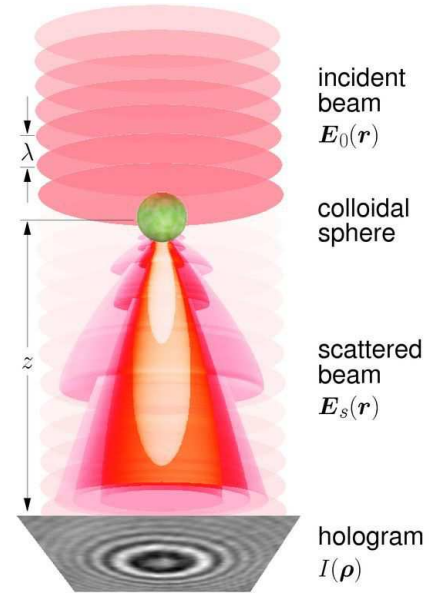}
   \caption{{\bf Principle of video holographic microscopy.}  A
     colloidal particle scatters a portion $\vec{E}_s(\vec{r})$
     of an initially collimated laser beam $\vec{E}_0(\vec{r})$.
     The scattered beam, here represented 
     by 5 calculated iso-amplitude surfaces, 
     interferes with the unscattered portion
     of the incident beam in the focal plane of a microscope objective,
     thereby forming an in-line hologram, $I(\vec{\rho})$.}
   \label{fig:dhm}
 \end{figure}
 Our holographic analysis instrument is based on a standard inverted
 optical microscope (Nikon TE-2000U), with a collimated and attenuated
 HeNe laser (Uniphase 5~\unit{mW}, $\lambda = 0.632~\micron$) 
 replacing the conventional incandescent illuminator and condenser.
 As indicated schematically in Fig.~\ref{fig:dhm},
 light scattered by a particle propagates
 to the microscope's focal plane, where it interferes with the
 undiffracted portion of the beam.
 The resulting interference pattern is magnified \cite{sheng06}
 by the microscope's
 objective lens (Nikon $100\times$ NA 1.4 oil immersion Plan-Apo)
 and video eyepiece ($1.5\times$) onto the sensor of a 
 grey-scale video camera (NEC TI-324AII).
 This system provides a total magnification of $135 \pm 1
 ~\unit{nm/pixel}$ over a $86 \times 65~\unit{\micron^2}$ field
 of view.
 Images are recorded as uncompressed digital video
 at 30~\unit{frames/s} using a commercial digital video recorder
 (Pioneer 520HS).

 Analyzing these digitized holograms yields the particle's
 three-dimensional position, $\vec{r}_p$, its radius, $a$, and
 its index of refraction, $n_p$.
 We assume that the incident field, 
 $\vec{E}_0(\vec{r}) = u_0(\vec{\rho}) \, \exp(i k z) \,
 \uvec{\epsilon}$,
 is uniformly polarized in the $\uvec{\epsilon}$ direction and
 varies slowly enough over the size of the particle to
 be treated as a plane wave propagating along the $\uvec{z}$ direction.
 Its amplitude $u_0(\vec{\rho})$ at position $\vec{\rho} = (x,y)$
 in the plane $z = z_p$ of the particle is thus the same as its
 amplitude in the focal plane, $z = 0$.
 The wave propagates along the $\uvec{z}$ direction with wave
 number $k = 2\pi n_m / \lambda$, where $\lambda$ is the light's
 wavelength in vacuum and $n_m$ is the refractive index of the
 medium.
 For pure water at $25^\circ\unit{C}$, $n_m = 1.3326$ at
 $\lambda = 0.632~\micron$.

 The particle at $\vec{r}_p$ scatters a portion of the
 incident field into a highly structured outgoing wave,
 $\vec{E}_s(\vec{r}) = \alpha \, \exp(-i k z_p) \, u_o(\vec{r}_p) \, 
 \vec{f}_s(\vec{r} - \vec{r}_p)$,
 where $\alpha \approx 1$ accounts for variations in
 the illumination, and where
 $\vec{f}_s(\vec{r})$ is the
 Lorenz-Mie scattering function \cite{bohren83,barber90,mishchenko02},
 which depends on $a$, $n_p$, $n_m$ and $\lambda$.
 The scattered field generally covers a large
 enough area at the focal plane that the interference pattern,
 \begin{equation}
   \label{eq:interference}
   I(\vec{\rho}) = \left. \abs{\vec{E}_s(\vec{r}) +
       \vec{E}_0(\vec{r})}^2 \right\vert_{z=0},
 \end{equation}
 is dominated by long-wavelength variations in
 $\abs{u_0(\vec{\rho})}^2$.
 The resulting distortions have been characterized \cite{pu03},
 but were not corrected in previous analyses of $I(\vec{\rho})$
 \cite{ray91,moreno00,pu03,denis06,park07}.
 Fortunately, $\abs{u_0(\vec{\rho})}^2$
 can be measured in an empty field of view, and the in-line hologram
 can be normalized to obtain the undistorted image
 \begin{align}
   \label{eq:image}
   B(\vec{\rho}) & \equiv 
   \frac{I(\vec{\rho})}{\abs{u_0(\vec{\rho})}^2} \\
   & = 1 + 
   \frac{2\, \Re\{\vec{E}_s(\vec{r}) \cdot \vec{E}_0^\ast(\vec{r})\}}{
       \abs{u_0(\vec{\rho})}^2}
   + \frac{\abs{\vec{E}_s(\vec{r})}^2}{\abs{u_0(\vec{\rho})}^2},
  \label{eq:nearlythere}
 \end{align}
 on the plane $z = 0$.
 If we further assume 
 that the phase of the collimated incident beam
 varies slowly over the field of view,
 the normalized image is related to the calculated Mie scattering
 pattern, $\vec{f}_s(\vec{r})$, in the plane $z = 0$ by
 \begin{equation}
   \label{eq:hologram}
   B(\vec{\rho}) \approx 1 + 
   2 \alpha \, \Re\left\{ 
     \vec{f}_s(\vec{r}-\vec{r}_p) \cdot \uvec{\epsilon} \, e^{- i k z_p}
     \right\} + 
     \alpha^2 \abs{\vec{f}_s(\vec{r}-\vec{r}_p)}^2.
 \end{equation}

 \begin{figure}[!t]
   \centering
   \includegraphics[width=\columnwidth]{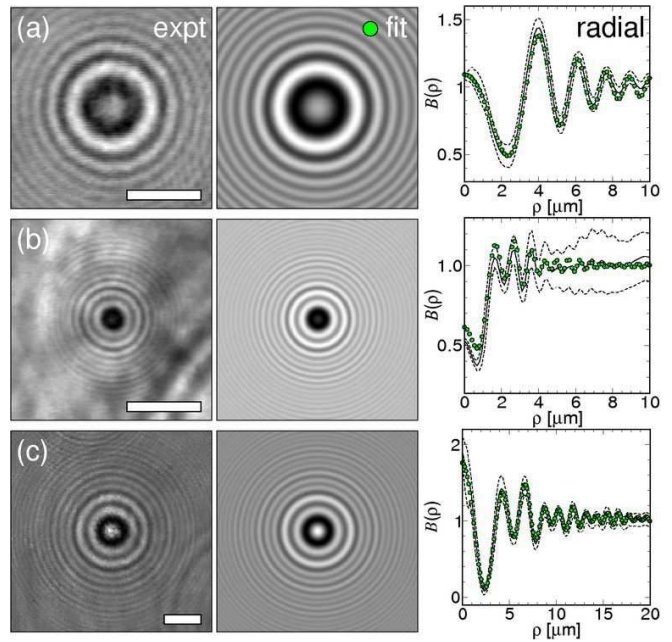}
   \caption{{\bf Fitting to normalized holograms.}
     (a) Normalized hologram $B(\vec{\rho})$, numerical fit to
     Eq.~(\ref{eq:hologram}), and azimuthally averaged radial
     profile $B(\rho)$ for a 1.43~\micron
     diameter polystyrene sphere in water
     at $z_p = 22.7~\micron$.  All scale bars indicate 10~\micron.
     Curves in the radial profile are obtained from experimental data,
     discrete points were obtained from the fit.
     (b) Data for a 1.45~\micron diameter TiO$_2$ sphere
     dispersed in immersion oil ($n_m = 1.515$) at $z_p = 7.0~\micron$
     (c) Data for a 4.5~\micron diameter SiO$_2$ sphere in water at
     $z_p = 38.8~\micron$.
   }
   \label{fig:pstio2}
 \end{figure}

 Equation~(\ref{eq:hologram}) can be fit to measured
 holograms by treating the particle's
 three-dimensional position, its radius and its refractive index
 as free parameters.
 Previous studies fit non-normalized holograms to
 phenomenological models
 \cite{thompson74,soontaranon04,pu05,denis06,guerreroviramontes06,park07}
 or Mie scattering theory \cite{alexandrov05}
 for some of these quantities, but never all five.
 Because errors in the adjustable parameters are strongly correlated,
 failing to optimize them all simultaneously 
 yields inaccurate results.
 Fitting instead to the full Lorenz-Mie theory
 \cite{bohren83,barber90,mishchenko02,wiscombe80,pu03,du04}
 provides more information with greater precision.

 Numerical fits to digitized and normalized 
 holographic images
 were performed with the Levenberg-Marquardt 
 nonlinear least-squares minimization algorithm 
 \cite{more80,more77,gill78} using the camera's
 measured signal-to-noise ratio to estimate single-pixel errors.
 The $\chi^2$ deviates
 for all of the fits we report
 are of order unity, so that the calculated uncertainties
 in the fit parameters accurately reflect their
 precision \cite{more80,gill78,dennis96}.
 These estimates incorporate the estimated
 covariance of the adjustable parameters, so that they
 also may be interpreted as the resolution of each parameter 
 \cite{dennis96}.

 Because the laser's wavelength and the medium's refractive
 index are both known, the only instrumental calibration
 is the overall magnification.
 This contrasts with other three-dimensional 
 particle tracking techniques 
 \cite{crocker96,pralle99,speidel03,sheng06,lee07,park07},
 which require independent calibrations for each type
 of particle, particularly to track
 particles in depth.

 The image in Fig.~\ref{fig:pstio2}(a) shows
 the normalized hologram, $B(\vec{\rho})$,
 for a polystyrene sulfate sphere 
 dispersed in water at height $z_p = 22.7~\micron$ above
 the focal plane.
 This sphere was obtained from a commercial sample with a nominal 
 diameter of $2a = 1.48 \pm 0.03~\micron$ (Bangs Labs, Lot PS04N/6064).
 The camera's electronic shutter was set for an exposure time
 of 0.25~\unit{msec} to minimize blurring due to Brownian motion
 \cite{savin05}.
 After normalizing the raw 8-bit digitized images, 
 each pixel contains roughly 5 significant bits of information.
 The numerical fit to $B(\vec{\rho})$ faithfully reproduces
 not just the position of the interference fringes, but also
 their magnitudes.
 The quality of the fit may be judged from the azimuthal
 average;
 the solid curve is an angular average about the
 center of $B(\vec{\rho})$, the dashed curves 
 indicate the standard deviations of the average,
 and the discrete points are obtained from the fit.

 \begin{figure*}[!t]
   \centering
   \includegraphics[width=0.75\textwidth]{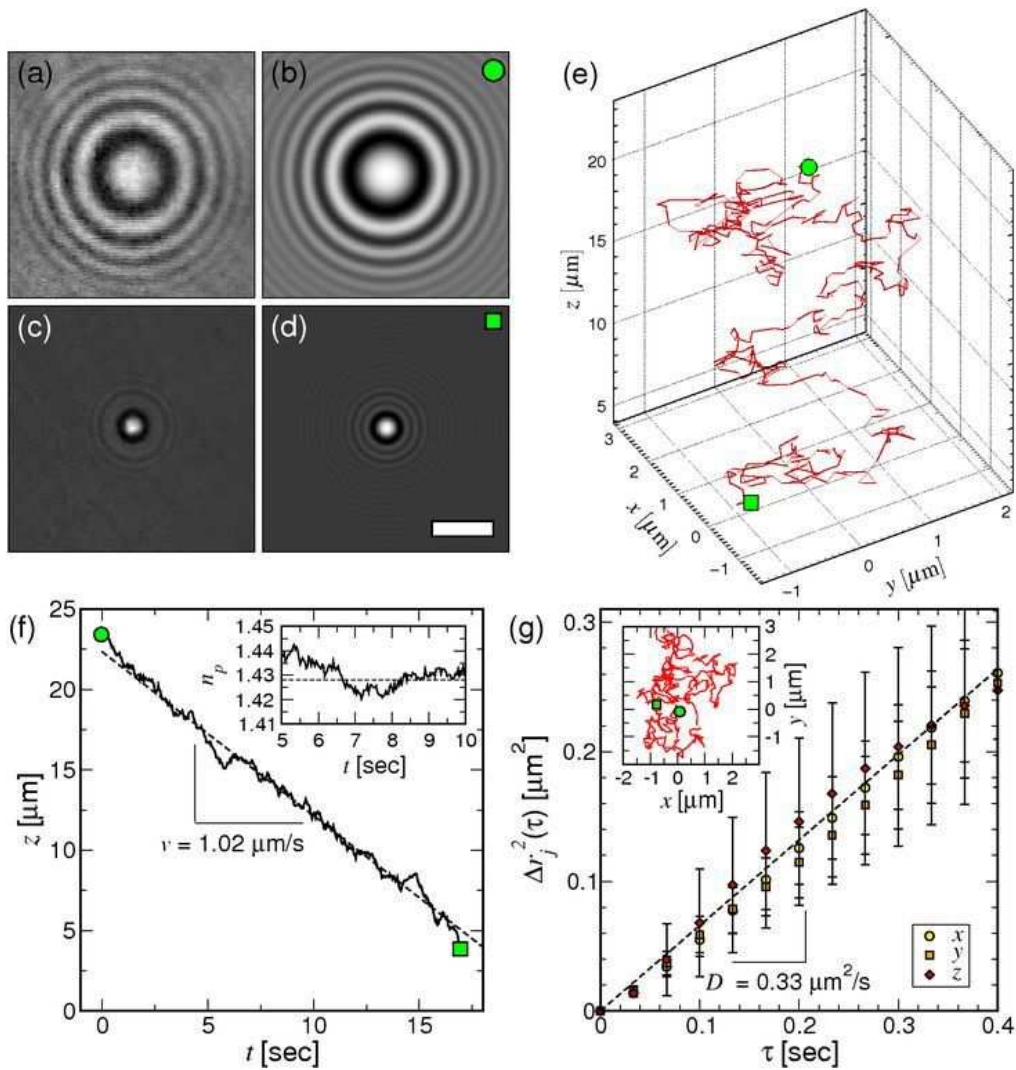}
   \caption{{\bf Holographic tracking of a sedimenting colloidal silica
     sphere.}
   (a) and (c) DHM images of the sphere at the beginning and end of its
   trajectory, respectively.  The scale bar indicates 5~\micron.
   (b) and (d) Fits to Eq.~(\ref{eq:hologram}).  
   (e) Three-dimensional trajectory with starting point (circle) 
   and end point (square) labeled.  (f) $z(t)$, showing thermal
   fluctuations about uniform sedimentation.  Inset: The fit refractive
   index is independent of position.  (g) Mean-square positional 
   fluctuations display Einstein-Smoluchowsky scaling in $x$, $y$
   and $z$.}
   \label{fig:sio2}
 \end{figure*}

 The fit value for the radius, $a = 0.73 \pm 0.01~\micron$, 
 falls in the sample's specified range,
 which reflected a lower bound
 of $0.69 \pm 0.07~\micron$ obtained
 with a Beckman Z2 Coulter Counter and an upper bound of
 $0.76 \pm 0.08~\micron$ obtained by analytical centrifugation.
 Agreement between the quoted and measured particle size suggests
 that the present measurement's accuracy is comparable to its precision.
 In that case, both
 precision and accuracy surpass results previously obtained \cite{denis06}
 through analysis of $I(\vec{\rho})$.
 The trajectory-averaged value for the 
 refractive index, $n_p = 1.55 \pm 0.03$,
 also is consistent with the properties of polystyrene colloid inferred
 from light scattering measurements on bulk dispersions
 \cite{ma03}.

 Comparable precision in measuring a single particle's
 refractive index has been achieved by analyzing a
 colloidal particle's dynamics in an optical trap \cite{knoner06}.
 This method only can be applied to particles with comparatively
 small refractive indexes, however, because particles with relative
 refractive indexes greater than $n_p \approx 1.3 \, n_m$ are difficult to trap.
 Holographic characterization, by contrast, requires only a single holographic
 snapshot rather than an extensive time series,
 does not require optical trapping, and so does not require
 separate calibration of the trap, and is effective over a wider
 range of particle sizes and refractive indexes.

 The corresponding data in Fig.~\ref{fig:pstio2}(b)
 were obtained for a 1.45~\micron diameter TiO$_2$ sphere
 at $z_p = 7~\micron$ above the focal plane.
 This sample was synthesized from titanium tetraethoxide and
 was heat-treated to increase its density \cite{eidenassmann04}.
 Strong forward scattering by such high-index particles
 gives rise to imaging artifacts unless the medium is index matched
 to the cover slip.
 Dispersing the particle in immersion oil ($n_m = 1.515$)
 eliminates these artifacts, but introduces spherical aberration
 for the lens we used,
 which must be corrected \cite{roichman05a} to obtain reliable
 results.
 The fit diameter of $1.45 \pm 0.03~\micron$ and refractive index
 of $2.01 \pm 0.05$ are consistent with results obtained by electron
 microscopy and bulk light scattering, respectively.
 This result is noteworthy because no other single-particle
 characterization method works for such high refractive indexes.

 The data in Fig.~\ref{fig:pstio2}(c) show results for a 
 nominally 5~\micron silica sphere (Bangs Labs, Lot SS05N/4364)
 dispersed in water at $z_p = 38.8~\micron$ above the focal plane.
 The fit refractive index, $n_p = 1.434 \pm 0.001$, is
 appropriate for porous silica and the diameter, 
 $a = 4.51 \pm 0.01~\micron$ 
 agrees with the $4.82 \pm 0.59~\micron$ value
 obtained for this sample with a Beckman Z2 Coulter Counter.

 The same fits resolve the particle's position with
 a precision of 1~\unit{nm} in-plane and 10~\unit{nm} along
 the optical axis.  
 This substantially improves upon the typical 10~\unit{nm} 
 in-plane accuracy
 obtained with standard particle tracking techniques
 with the same microscope and camera \cite{crocker96}.
 The difference can be ascribed to the larger number of pixels
 subtended by a holographic image, and to the images'
 strong intensity gradients, which constrain the fits.
 The estimated 10~\unit{nm} axial resolution 
 surpasses results obtained
 with morphometric axial particle
 tracking \cite{crocker96,sheng06,lee07} by a factor of ten.

 Nanometer-scale tracking resolution can be obtained 
 under conventional illumination, but requires detailed
 calibrations for each particle \cite{gosse02}.
 Still better in-plane spatial resolution can be obtained
 at much higher bandwidths through back-focal-plane 
 interferometric methods \cite{gittes98}, but also require
 accurate calibrations with piezo translators.
 Total internal reflection microscopy (TIRM) similarly offers
 sub-nanometer axial resolution \cite{brown90,prieve90}, but
 performs no better than conventional imaging methods for
 in-plane tracking.

 An additional benefit of holographic imaging over other
 particle-tracking techniques is its very large depth of focus.
 Our system provides useful data over a range of more than
 100~\micron, which
 contrasts with the $\pm 3~\micron$ useful depth of focus
 using conventional illumination \cite{gosse02} and the
 100~\unit{nm} range of TIRM \cite{brown90,prieve90}.

 Holographic video microscopy lends itself to three-dimensional
 particle tracking, as
 the data in Fig.~\ref{fig:sio2} demonstrate
 for a colloidal silica sphere (Bangs Labs, Lot SS04N/5252)
 dispersed in water.
 This particle was lifted 30~\micron above the focal plane
 with an optical tweezer, and then released and allowed to sediment.
 The images in Fig.~\ref{fig:sio2}(a) and (c) show the particle 
 near the beginning of its trajectory and near the end.  
 Fits to Eq.~(\ref{eq:hologram})
 are shown in Figs.~\ref{fig:sio2}(b) and (d).

 The particle's measured trajectory in 1/30~\unit{s} intervals
 during 15~\unit{s} of its descent is plotted in
 Fig.~\ref{fig:sio2}(e).
 Its vertical position $z(t)$, 
 Fig.~\ref{fig:sio2}(f),
 displays fluctuations about a uniform sedimentation speed,
 $v = 1.021 \pm 0.005~\unit{\micron/s}$.
 This provides an estimate
 for the particle's density through
 $\rho_p = \rho_m + 9 \eta v / (2 a^2 g)$,
 where $\rho_m = 0.997~\unit{g/cm^3}$ is the density of water and
 $\eta = 0.0105~\unit{P}$ is its viscosity at 
 $T = 21\unit{^\circ C}$, and where $g = 9.8~\unit{m/s^2}$ is
 the acceleration due to gravity.
 The fit value for the particle's radius,
 at $a = 0.729 \pm 0.012~\unit{\micron}$,
 remained constant as the particle settled.
 This value is consistent with the manufacturer's specified
 radius of $0.76 \pm 0.04~\micron$, measured with a Beckman
 Z2 Coulter Counter.
 Accordingly, we obtain $\rho_p = 1.92 \pm 0.02~\unit{g/cm^3}$,
 which is a few percent smaller than the manufacturer's rating for
 the sample.
 However, the fit value for the refractive index, 
 $n_p = 1.430 \pm 0.007$, also is 1.5\% below the rated value,
 suggesting that the particle is indeed less dense
 than specified.

 The mean-square displacements,
 $\Delta r_j^2(\tau) = \avg{(r_j(t + \tau) - r_j(t))^2}$,
 of the components of the particle's position
 provide additional consistency checks.
 As the data in Fig.~\ref{fig:sio2}(g) show, fluctuations
 in the trajectory's individual Cartesian components agree with each other,
 and all three display linear Einstein-Smoluchowsky scaling,
 $\Delta r_j^2(\tau) = 2 D \tau$, with a diffusion coefficient
 $D = 0.33 \pm 0.03~\unit{\micron^2/s}$.  This is consistent
 with the anticipated Stokes-Einstein value,
 $D_0 = k_B T / (6 \pi \eta a) = 0.30 \pm 0.02~\unit{\micron^2/s}$,
 where $k_B$ is Boltzmann's constant.
 Using the methods of Ref.~\cite{savin05}, we
 then interpret the offsets obtained from linear fits to $\Delta r^2_j(t)$
 to be consistent with no worse than 1~\unit{nm}
 accuracy for in-plane positions and 10~\unit{nm} for
 axial positions throughout the trajectory.
 The optical characterization of the particle's properties
 thus is consistent with the particle's measured dynamics.

 We have successfully
 applied holographic characterization to colloidal spheres as small
 as 100~\unit{nm} in diameter and as large as 10~\micron.
 Unlike model-based analytical methods, fitting to the exact
 Lorenz-Mie scattering theory is robust and reliable over a
 far wider range of particle sizes, provided that care is
 taken to maintain numerical stability in calculating $\vec{f}_s(\vec{r})$ 
 \cite{bohren83,wiscombe80,du04}.
 Such numerical implementations have been reported for particles
 as small as a few nanometers and as large as a few millimeters,
 with relative refractive index ratios from less than $m = 0.1$ to
 over 10, and with large imaginary refractive indexes.
 In all cases, the instrumental magnification and field of view 
 must be selected to fit the sample.

 The principal limitations of the
 six-parameter model in Eq.~(\ref{eq:hologram}) are the assumptions
 that the scatterer is homogeneous and isotropic, and that its
 interface is sharp.
 These assumptions can be relaxed at the cost of increased
 complexity and reduced numerical robustness.
 For example, analytical results are available for
 core-shell particles \cite{bohren83},
 and for particles with more complex shapes 
 \cite{bohren83,mishchenko02},
 such as ellipsoids, spherical clusters and cylindrical nanowires.
 All such elaborations involve additional adjustable parameters
 and thus are likely to pose computational challenges.

 We have demonstrated that a single snapshot from
 an in-line holographic microscope
 can be used to measure a colloidal sphere's position and size 
 with nanometer-scale resolution,
 and its refractive index with precision
 typically surpassing 1 percent.

 A video stream of such images therefore
 constitutes a powerful six-dimensional microscopy for soft-matter
 and biological systems.
 Holographic particle tracking is ideal for three-dimensional
 microrheology, for measuring colloidal interactions
 and as force probes for biophysics.
 The methods we have described can be applied to tracking
 large numbers of particles in the field of view simultaneously
 for highly parallel measurements.
 Real-time single-particle characterization and tracking
 of large particle ensembles will be invaluable in
 such applications as holographic assembly of photonic devices
 \cite{sinclair04,roichman05}.
 Applied to more highly structured samples such as biological cells
 and colloidal heterostructures,
 they could be used as a basis for cytometric analysis or combinatorial
 synthesis \cite{johnston06}.

 This work was supported by the National Science Foundation under
 Grant Number DMR-0606415.
 SHL acknowledges support of the Kessler Family Foundation.
 GRY was supported by KBSI grant (N27073).
 KSH and SMY have been supported by the NCRI Center
 for Integrated Optofluidic Systems of MOST/KOSEF.
 We are grateful to Ahmet Demir\"ors for synthesizing 
 the 1.4~\micron diameter
 TiO$_2$ particles.


\end{document}